\documentclass[10pt,twocolumn,letterpaper]{article}

\usepackage{iccv}
\usepackage{times}
\usepackage{epsfig}
\usepackage{graphicx}
\usepackage{amsmath}
\usepackage{amssymb}

\usepackage{subcaption, multirow, xcolor}
\usepackage[pagebackref=true,breaklinks=true,letterpaper=true,colorlinks,bookmarks=false]{hyperref}

\iccvfinalcopy % *** Uncomment this line for the final submission

\def\iccvPaperID{5491} % *** Enter the ICCV Paper ID here

\ificcvfinal\fi
\begin{document}

%%%%%%%%% TITLE
\title{Joint High Dynamic Range Imaging and Super-Resolution from a Single Image}

\author{Jae Woong Soh$^{1}$ \qquad Jae Sung Park$^{2}$ \qquad Nam Ik Cho$^{1}$\\
Department of ECE, INMC, Seoul National University, Seoul, Korea$^{1}$ \\
Samsung Electronics Co. Ltd., Suwon, Korea$^{2}$\\
{\tt\small soh90815@ispl.snu.ac.kr, jason79.park@samsung.com, nicho@snu.ac.kr}
% For a paper whose authors are all at the same institution,
% omit the following lines up until the closing ``}''.
% Additional authors and addresses can be added with ``\and'',
% just like the second author.
% To save space, use either the email address or home page, not both
}

\maketitle
\thispagestyle{empty}

%%%%%%%%% ABSTRACT
\begin{abstract}
This paper presents a new framework for jointly enhancing the resolution and the dynamic range of an image, i.e., simultaneous super-resolution (SR) and high dynamic range imaging (HDRI), based on a convolutional neural network (CNN). From the common trends of both tasks, we train a CNN for the joint HDRI and SR by focusing on the reconstruction of high-frequency details. Specifically, the high-frequency component in our work is the reflectance component according to the Retinex-based image decomposition, and only the reflectance component is manipulated by the CNN while another component (illumination) is processed in a conventional way. In training the CNN, we devise an appropriate loss function that contributes to the naturalness quality of resulting images. Experiments show that our algorithm outperforms the cascade implementation of CNN-based SR and HDRI.
\end{abstract}

\begin{figure}[t]
	\begin{center}
		\begin{subfigure}[t]{0.95\linewidth}
			\centering
			\includegraphics[width=1\columnwidth]{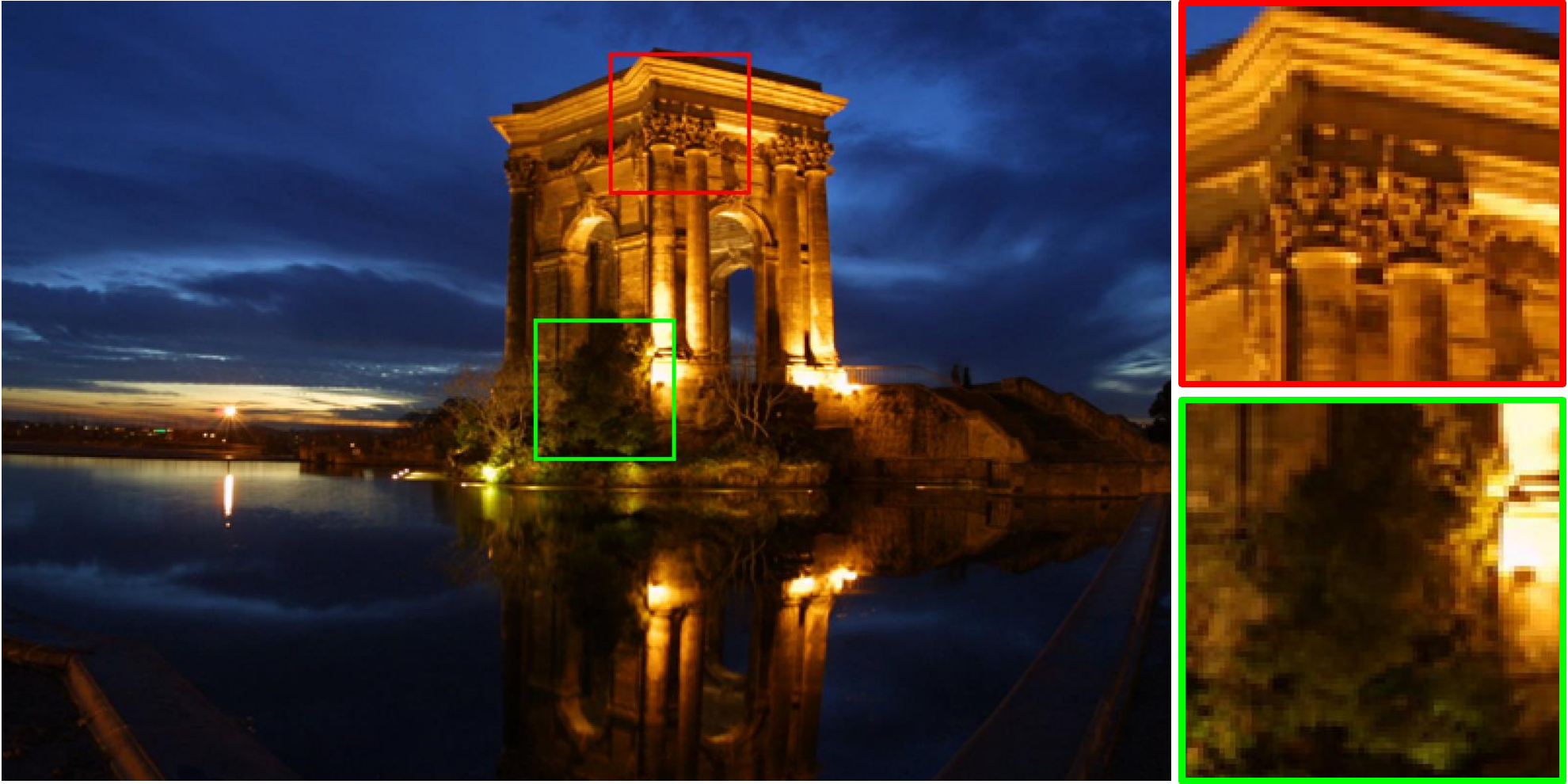}
			\caption{The input LDR-LR image.}
		\end{subfigure}
		\begin{subfigure}[t]{0.95\linewidth}
			\centering
			\includegraphics[width=1\columnwidth]{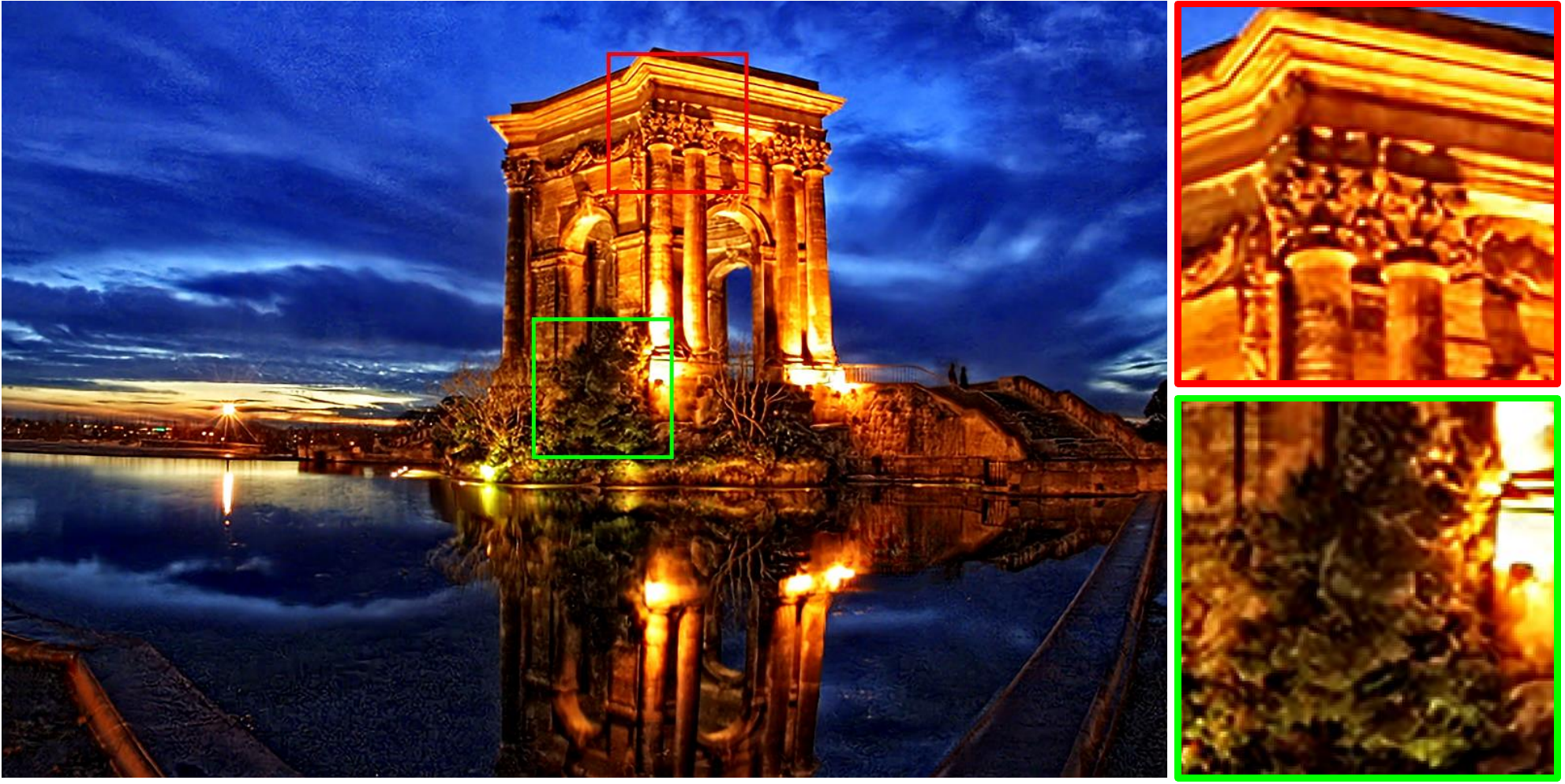}
			\caption{Our HDRI-SR-B result.}
		\end{subfigure}
	\end{center}
	\caption{An example of LDR-LR input and the output of our method. It can be seen that both the dynamic range and resolution are enhanced.}
	\label{fig:001}
\end{figure}

%%%%%%%%% BODY TEXT
\section{Introduction}

With the advent of ultra high definition television (UHDTV) with HDR rendering \cite{uhdtv}, the techniques for capturing high resolution (HR) and high dynamic range (HDR) images have become important. In addition to developing the methods to capture the new UHD contents, it is also important to convert the vast amount of existing low-resolution (LR) and low dynamic range (LDR) images to the HDR-HR contents for rendering them on the UHDTV. 
For these purposes, there have been many methods to convert the LR to HR, which is called the single image SR (SISR). Also, there have been some single image HDRI algorithms to produce an HDR image from a single LDR input.

Since the SISR is an important problem that finds many useful applications, it has long been studied by many researchers \cite{BenchSR, Prior1, SparseSR, Self-Exemplar, A+, SRCNN, SRGAN, EDSR, RDN}. Some earlier works exploited the statistical priors of images for the SISR \cite{Prior1,Prior2}. The learning based methods, specifically the ones based on the neighbor embedding \cite{EmbedSR} and sparse coding methods \cite{SparseSR,CoupleSR,A+} were also introduced for better SR. Recently, the state-of-the-art methods are based on the CNN \cite{SRCNN, VDSR, SRGAN, LapSRN, SRDenseNet, EDSR, RDN, IDN, DSRN}, which show further enhanced results than the previous ones. Generative methods \cite{SRGAN, EnhanceNet, SFT-GAN} based on generative adversarial networks (GAN) \cite{GAN} have also been introduced for the better perceptual quality of SR images.

For the HDRI, multiple images or sensors with different exposures are used, or a single image is reversely tonemapped to generate an HDR output \cite{MEHDR1,MEHDR2,MEHDR3, rTMO1, Kovaleski, Masia, HDRICNN, DeepHDRDS, DeepHDRI}. For the enhancement of undesirably illuminated images, some methods generated virtual exposure images from a single input and applied the conventional HDRI process \cite{VEF,VEF2,Wang,fbEM,LIME,Retinex1,Retinex2,HDRSR2}. Recently, deep CNN-based methods have also been proposed \cite{HDRICNN,DeepHDRDS,DeepHDRI}.
There are also some works that use both HDRI and SR for image enhancement:
Schubert \etal \cite{HDRSR1} developed a method for the HDRI-SR which consists of several steps. Park \etal tried several cascade implementations of HDRI-SR in different color spaces \cite{HDRSR2}.

Recently, deep CNNs have shown dramatic improvements in most of low to high-level vision tasks including super-resolution \cite{SRCNN,VDSR,SRGAN,DRRN,LapSRN,EDSR,RDN,DSRN}, Gaussian denoising \cite{DnCNN}, and JPEG artifact reduction \cite{ARCNN,DDCN}. For these problems, the deep CNN is shown to work as a proper mapping function from the degraded image to the original. In this respect, we adopt the CNN for the joint HDRI-SR problem. 

The most straightforward method of using a CNN is to design an end-to-end network, \ie, a deep network which takes LDR-LR image as the input and generates the corresponding HDR-HR. However, in the case of joint HDRI-SR task, we have a problem that there is no standardized labeled data. Precisely, the HDR images' luminance range and tonemapping function from the HDR to the LDR are not well specified in the current HDR datasets. Moreover, the dynamic ranges of target images are usually different from each other, and the tonemapping functions are also different and nonlinear due to the use of locally adaptive nonlinear mapping in most images. Thus, by directly training a discriminative CNN to map LDR images to HDR ones, the network usually fails to converge. Hence, we need to use a transformed image or find another domain that is less affected by the luminance range and tonemapping function. 
%In this respect, we may consider the reflectance component as the input to the CNN.
%However, especially to apply supervised learning scheme for HDRI, the main problem arises that there is no ground-truth for HDR image. In particular, there is no standardized labeled data. Each HDR dataset involves in such characteristics as HDRI method for generating ground-truth HDR image, the range of luminance, or tonemapping method. Thus, by directly training discriminative CNN to map LDR image to HDR image, the network mostly fails to learn proper generalized function rather overfit to the train data or fail to converge which deteriorates overall performance.

Some of the previous works also showed that it is important to find an appropriate domain when applying a CNN for image enhancement. For example, it is shown in \cite{DDCN} that applying the dual domain representation, \ie, using the image and DCT domain priors increases the performance of JPEG artifacts reduction compared to the methods without the DCT prior. Also, the SR with wavelet domain priors \cite{waveletSR} enhanced the performance compared to the conventional image domain methods. Additionally, recent SR and denoising CNNs such as VDSR \cite{VDSR} and DnCNN \cite{DnCNN} focus on the residual structure, because the SR task is to find the high-frequency details and the Gaussian denoising is also to estimate the noise which is the residual signal. 
%In summary, the insights above lead us to adopt proper domain for both tasks which is to focus on the image structure in the high-frequency components for achieving the better performance and the better convergence.

The HDRI also focuses on the reconstruction of image details rather than the low-frequency components. Specifically, recent single image HDRI methods process the illumination and reflectance components separately \cite{fbEM,LIME,Retinex1,Retinex2,HDRSR2}, where the illumination corresponds to the low-frequency component and the reflectance amounts to the image details. The illumination is simply scaled according to certain virtual exposure levels, while the image details are locally and sophisticatedly manipulated to reveal the details in the saturated regions. 

Considering that both HDRI and SR try to find the lost image details in the high-frequency region rather than the low-frequency, we design our joint HDRI-SR CNN to focus on the high-frequencies.
%We tackle the joint task by reconstructing plausible high-frequency textures and details lost due to saturation (for HDRI) and the limited resolution of capturing devices (for SR).
%Furthermore, it is possible to address the problem of HDRI which is the absence of labeled data.
Furthermore, it is possible to address the above-stated problem of HDRI training datasets by excluding the low-frequency illumination component in the training and inference, because the illuminations usually have very wide and different ranges from each other. Specifically, we propose a CNN architecture and its training schemes for enlarging the resolution and dynamic range of reflectance component. We also adopt the generative scheme \cite{GAN}, for generating better details and textures. 

In summary, the main contributions of this paper are as follows.
\begin{itemize}
%\item We propose a framework for joint HDRI and SR from a single input image.
%\item From the knowledge obtained from the previous HDRI and SR researches, we derive a method that processes the reflectance component of an image by the CNN and the illumination by a conventional approach.
\item The proposed method performs HDRI and SR using a single CNN with high generalization performance, especially without well-organized labeled datasets for HDRI.
\item The proposed method performs better than adopting separate CNNs for HDRI and SR in terms of perceptual quality and some no-reference metrics.
\item The proposed single CNN requires much fewer parameters than the cascade of state-of-the-art SR and HDRI networks.  
	\item Unlike the conventional single image HDRI that needs to generate virtual multi-exposure images, the proposed method directly produces the HDR image through the CNN which simultaneously performs the SR.
\end{itemize}

\begin{figure*}[!t]
	\centering
	\includegraphics[width=0.98\linewidth]{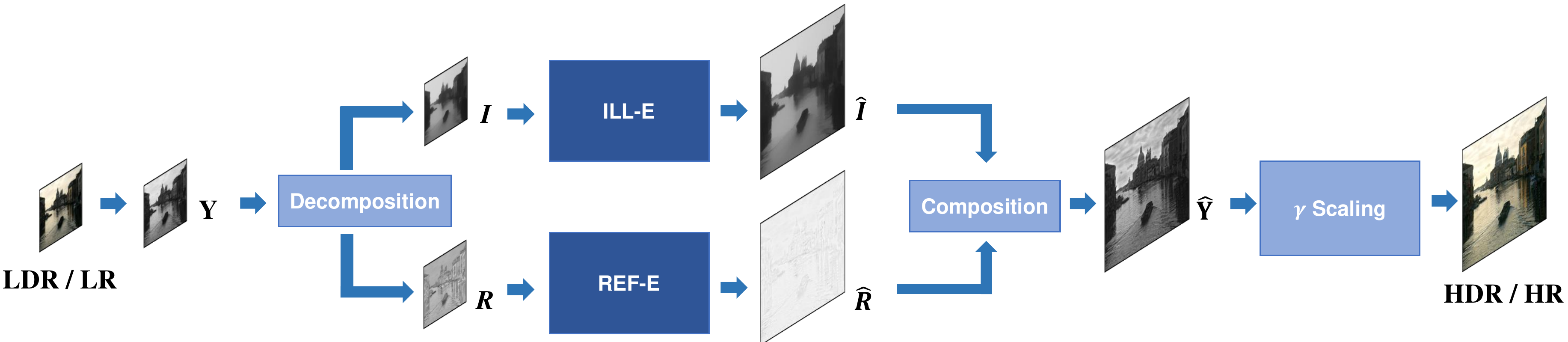}
	\caption{The overall process of the proposed scheme. We first decomposed the LDR-LR input into the illumination $I$ and the reflectance $R$. ILL-E and REF-E denote ILLumination Enhancement and REFlectance Enhancement, respectively. The CNN is designed and trained only for the reflectance component $R$, and the illumination component $I$ is simply up-scaled to increase the dynamic range.}
	\label{fig:overall}
\end{figure*}

\section{Related Work}

\subsection {Single Image Super-Resolution based on CNN}
In recent years, the CNN-based SISR algorithms outperform most of the conventional non-CNN based methods. Since Dong \etal first proposed a CNN for the SISR \cite{SRCNN}, many other deep networks have been proposed. For some examples, the residual structure is used for better performance in \cite{VDSR}, and the sub-pixel convolution layer is introduced in \cite{ESPCN} to speed up. The SRGAN proposed in \cite{SRGAN} generates the photorealistic SR images by exploiting the generative adversarial losses. Guo \etal \cite{waveletSR} proposed a CNN for the SR in the wavelet domain, and have shown that Haar wavelet domain is an efficient one for the SISR.
There are also many other structures and methods such as the recursive
architecture \cite{DRCN, DRRN} and the Laplacian pyramids \cite{LapSRN}. Very recently, enormously deep structures have been proposed and achieved the state-of-the-art performances \cite{EDSR,RDN}.

\subsection{High Dynamic Range Imaging from a Single Image}
For generating an HDR image from a single input, most of the conventional methods generate the virtual multi-exposure images by applying brightness enhancement functions and then fuse them with the appropriate weight maps obtained from each of the virtual images \cite{VEF,VEF2,fbEM,LIME,Retinex1,Retinex2,HDRSR2}. Among these techniques, the Retinex based approaches \cite{fbEM,LIME,Retinex1,Retinex2,HDRSR2} enhance the illumination and reflectance components separately. The undesirably illuminated regions are compensated by the estimated illumination, while the saturated details are enhanced by controlling the reflectance component. For generating a real (not the exposure fusion in low dynamic range) HDR image, the reverse tone mapping operators (rTMOs) with well-designed expanding maps are presented in \cite{rTMO1,rTMO2,Kovaleski,Masia,Masia2}. Recently, a few CNN-based approaches \cite{HDRICNN,DeepHDRDS,DeepHDRI} have been proposed. In \cite{DeepHDRDS}, a CNN-based multi-exposure fusion scheme is proposed. In \cite{DeepHDRI}, a CNN-based reverse tone mapping method is proposed where virtual multi-exposure images are predicted by a CNN, and then they are fused for generating the HDR image. In \cite{DeepHDRDS, DeepHDRI}, the HDR image is generated as a weighted sum of multi-exposure images (real or virtually generated). In \cite{HDRICNN}, the interest is on the saturated highlights and the CNN predicts these saturated regions to generate an HDR image.

\section{Proposed HDRI-SR Scheme}
The overview of our method is illustrated in
\figurename{~\ref{fig:overall}}, which shows the process of only $Y$ component because
the $CbCr$ components are just bicubic interpolated (not shown in the figure). 
The figure shows that the $Y$ is decomposed into the illumination ($I$) and reflectance ($R$), which undergo different processes and finally fused again to be an enhanced $\hat{Y}$.

For explaining the process formally in the rest of paper, we denote the HDR-HR image as $X_{HH}$, HDR-LR as $X_{HL}$, LDR-HR as $X_{LH}$, and LDR-LR as $X_{LL}$, where $X$ is an RGB image. Also, $\{Y, Cb, Cr, I, R\}$  with one of these subscripts mean the corresponding components of $X$ with the same subscript. Through the manipulation of these components, the final goal is to find a plausible $X_{HH}$ from the given $X_{LL}$.

\subsection{Image Decomposition}

Image decomposition process has an important role in our scheme because it enables to exploit domain properties and to train CNN without consistent HDR ground-truth.

\begin{figure}[!t]
	\centering
	\includegraphics[width=0.98\columnwidth]{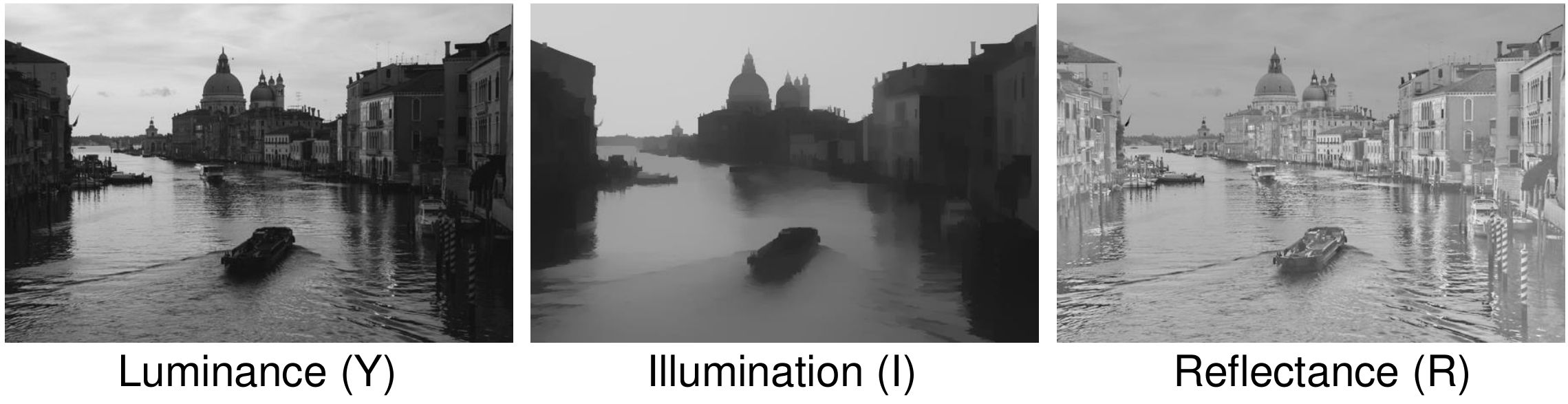}
	\caption{Visualization of luminance map, estimated illumination, and reflectance component obtained by WLS filtering.}
	\label{fig:WLS}
\end{figure}

The reflectance is obtained as the difference between the luminance and the estimated illumination as
\begin{equation}
R=\log(Y)-\log(I).
\end{equation}
The illumination $I$ is estimated by the filtering of $Y$ with a filter $G$, \ie, $I=Y*G(\cdot)$, where $G$ is usually a normalized Gaussian in the conventional work.
However, since it is known that using the Gaussian filter often makes the halo artifacts, we adopt the weighted least square (WLS) filter instead of the Gaussian. The WLS shows less halo artifacts because it preserves the edges better than the Gaussian filter. 
Precisely, the output image is obtained by solving an optimization problem, which is seeking the minimum of
\begin{equation}
\sum_{p} ((u_p-g_p)^2 + \lambda (a_{x,p}(g)|\nabla_x u_p|^2 + a_{y,p}(g)|\nabla_y u_p|^2)),
\end{equation}
where $p$ denotes the pixel location, $g$ and $u$ indicate the input and output respectively. Also, the smoothness weights are defined as
\begin{equation}
a_{x,p}(g)= (|\nabla_x l_p |^\alpha + \epsilon)^{-1}, \quad a_{y,p}(g)= (|\nabla_y l_p |^\alpha + \epsilon)^{-1},
\end{equation}
where $l$ is the log of $g$, $\alpha$ is the parameter to control the sensitivity of the gradients of $g$, and $\epsilon$ is a small number to prevent the divide by zero. For all the training and test images, we set the parameters as $\lambda=2$, $\alpha=2$, and $\epsilon=0.0001$. We also use linearized luminance values for decomposition.
\figurename{~\ref{fig:WLS}} visualizes an example of luminance, estimated illumination, and reflectance obtained by the WLS filter, where we can see that the barely seen details in $Y$ are revealed in $R$.

\begin{figure*}[t]
	\centering
	\includegraphics[width=0.98\linewidth]{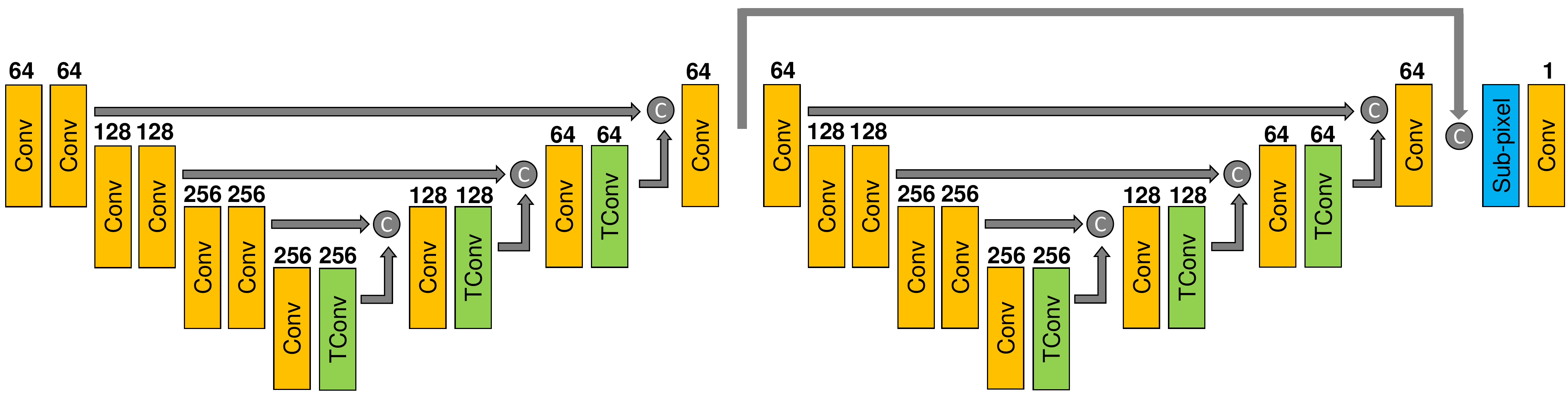}
	\caption{The overall CNN architecture of REF-Net. A stack of two hourglasses is adopted. The output of the first U-Net structure is skipped and concatenated with the second U-Net output, and then finally up-sampled via sub-pixel convolution layers.}
	\label{fig:REF-Net}
\end{figure*}

\subsection{Illumination Enhancement}
%For the illumination enhancement, we first bicubic interpolate the $I_{LL}$ to $I_{LH}$, and then stretch the dynamic range of $I_{LH}$ to be $I_{HH}$.
%For directly generating $I_{HH}$ from $I_{LL}$, we may use the CNN that is trained for the HDRI-SR of $R$, or we may design another CNN. However, using the CNNs do not improve the overall performance compared to the interpolation and stretching according to our experiments, because the illumination component is a smoothed image which contains little information.
%
%To be specific with the stretching, we linearly increase the illumination values, and then apply a simple gamma function to compensate for the non-linearity.
%In summary, the ILL-E in \figurename{~\ref{fig:overall}} is just the bicubic interpolation followed by linear stretching.

For the illumination enhancement, we first bicubicly interpolate the $I_{LL}$ to $I_{LH}$, and then simply compensate non-linearity to generate $I_{HH}$ from $I_{HH}$.
For directly generating $I_{HH}$ from $I_{LL}$, we may use the CNN that is trained for the HDRI-SR of $R$, or we may design another CNN. However, using the CNNs do not improve the overall performance compared to the interpolation and stretching according to our experiments, because the illumination component is a smoothed image which contains little information.

To be specific with the compensation, we first bicubicly interpolate the illumination values, then apply a simple gamma function $f(x)=x^{1/2.2}$ to compensate for the non-linearity.
In summary, the ILL-E in \figurename{~\ref{fig:overall}} is just the bicubic interpolation followed by gamma function.

\subsection{Reflectance Enhancement}
In this subsection, we present a CNN that maps $R_{LL}$ to $R_{HH}$, which is named REF-Net.
The overall architecture of the proposed REF-Net is shown in \figurename{~\ref{fig:REF-Net}} which has the stacked hourglass structure. Although U-Net \cite{U-Net} is shown to be effective in semantic segmentation, it may not be satisfiable for the prediction of reflectance components that have abundant textures. Hence, inspired by the work of successive stack of hourglasses \cite{Stack1, Stack2, Stack3}, we stack two U-Nets for better prediction. For details, the size of all convolution layers are $3\times3$ and transposed convolution layers are $4\times4$ for REF-Net, respectively. 
%Using the size of $4\times4$ for the transposed convolution is determined based on the observation that $4\times4$ gives less halo artifacts than $3\times 3$ as shown in {\em supplementary material}.

In manipulating the $R$ component, we should note that
the reflectance is the log ratio between luminance and illumination, which may ranges from $-\infty$ to $+\infty$ and hence not a suitable input to the CNN. It is found in our experiments that our CNN fails to predict high-quality reflectance when we feed $R$ to the CNN without any preprocessing. To stabilize the CNN and to address this issue, we employ the tangent-hyperbolic function as a preprocessing step, which bounds the input to the CNN into $(-1, 1)$. 
%Also, we empirically found that most of reflectance components lie in the range $[-1,1]$, hence $\tanh$ function is a good solution. 
In summary, our REF-Net predicts the $\tanh$ reflectance of HDR-HR for the given $\tanh$ reflectance of LDR-LR, which can be described as
\begin{align}
\tanh(\hat{R}_{HH})&=f_{R}(\tanh(R_{LL}); \theta_{R}),\\
\hat{R}_{HH}&=\tanh^{-1}(f_{R}(\tanh(R_{LL}); \theta_{R})),
\end{align}
where $f_R(\cdot)$ is our REF-Net with the parameter $\theta_{R}$.

\subsection{HDRI-SR Prediction}
With the $\hat{I}_{HH}$ and $\hat{R}_{HH}$, we recombine both components to build enhanced luminance $\hat{Y}_{HH}$ as
\begin{equation}
\hat{Y}_{HH}=\hat{I}_{HH} \odot \exp({\hat{R}_{HH}}),
\end{equation}
where $\odot$ denotes element-wise multiplication.
Then the final HDR irradiance map is obtained by
\begin{align}
\hat{X}_{HH}=(\textit{ycbcr2rgb}( [\hat{Y}_{HH}, Cb_{HH}, Cr_{HH}]))^\gamma,
\end{align}
where $Cb_{HH}$ and $Cr_{HH}$ are bicubic interpolation of $Cb_{LL}$ and $Cr_{LL}$. 

Finally, to display the $\hat{X}_{HH}$ to an HDR display, it can be linearly stretched to the luminance value of the target HDR display. On the other hand, to display the $\hat{X}_{HH}$ to an LDR display, the HDR irradiance map is tonemapped to be an enhanced LDR image \cite{Reinhard}.

\section{Loss Function}

In this section, we introduce loss functions that we use for our REF-Net.

\paragraph{Reconstruction Loss}
We adopt the reconstruction loss term as mean absolute error (MAE):
\begin{equation}
\mathcal{L}_{recon}=\frac{1}{N}\sum_{i}||R^i_{HH}-f_{R}(R^i_{LL};\theta_{R})||_1,
\label{eq:basic_model}
\end{equation}
where $i$ is the image index and $N$ denotes the batch size.

\paragraph{Adversarial Loss}
For better sharpness and details, we adopt adversarial loss inspired by recent successful generative super-resolution models \cite{SRGAN, EnhanceNet, SFT-GAN}.
The generative scheme not only generates better sharpness and details but also enables to predict saturated regions such as washed-out areas and diminished dark pixels based on training dataset.
We adopt recently proposed adversarial loss with relativistic discriminator \cite{RaGAN}, which shows great image quality with relativistic average standard GAN \cite{GAN}, named RaGAN. Specifically, the RaGAN loss for our scheme is described as:
\begin{align}
\mathcal{L}_{G}&= - \mathbb{E}_{x_r\sim\mathbb{P}_{r}}[\log(\tilde{D}(x_r))]-\mathbb{E}_{x_f\sim\mathbb{P}_{g}}[\log(1-\tilde{D}(x_f))]\\
\mathcal{L}_{D}&= - \mathbb{E}_{x_f\sim\mathbb{P}_{g}}[\log(\tilde{D}(x_f))]-\mathbb{E}_{x_r\sim\mathbb{P}_{r}}[\log(1-\tilde{D}(x_r))],
\end{align}
where $\mathbb{P}_r$ and $\mathbb{P}_g$ are empirical distributions of  ${R}_{HH}$ and $\hat{R}_{HH}$ respectively, $x_r$ and $x_f$ stand for real and generated data respectively, and
\begin{align}
\tilde{D}(x_r)=\text{sigmoid}(C(x_r)-\mathbb{E}_{x_f\sim\mathbb{P}_g}[C(x_f)])\\
\tilde{D}(x_f)=\text{sigmoid}(C(x_f)-\mathbb{E}_{x_r\sim\mathbb{P}_r}[C(x_r)])
\end{align}
where $C(\cdot)$ denotes the output logit of discriminator. We show the discriminator architecture for the GAN training in the \textit{supplementary material}.

\paragraph{Overall Loss}
We present two models: one is the basic model (HDRI-SR-B) without adversarial loss and the other is the complex model (HDRI-SR-C) with the adversarial loss.
Formally, the overall loss for the basic model is $\mathcal{L}_{recon}$ in eq.~(\ref{eq:basic_model}), and the loss for the complex model is
\begin{equation}
\mathcal{L}=\mathcal{L}_{recon} + \mu \mathcal{L}_{G}.
\end{equation}
where we set $\mu=10^{-3}$ for the training.

%\section{Discriminator Architecture}
%
%\begin{figure}[t]
%	\begin{center}
%		\centering
%		\includegraphics[width=1\columnwidth]{img/DIS.pdf}
%	\end{center}
%	\caption{The discriminator architecture for training HDRI-SR-C model. On each of the convolution layers, the $k, n,$ and $s$ denote kernel size, the number of features, and stride of convolution operation, respectively.}
%	\label{fig:DIS}
%\end{figure}
%
%\figurename{~\ref{fig:DIS} shows the discriminator architecture of our HDRI-SR-C model for adversarial training. We stack several convolution layers and employ global average pooling at the latter part of the network instead of fully-connected layers. Also, we alternate max-pooling operations by 2-strided convolution layers to decrease the spatial scale of features.
	
\section{Training Strategies}
For training the overall HDRI-SR, we use the MMPSG \cite{MMSPG} dataset which consists of multi-exposure images and the HDR images constructed from the multi-exposures. Note that the input to the network is the LDR-LR and the output is the corresponding HDR-HR reflectance for the training. For constructing such set of input-output, we select 40 sets of multi-exposure images, and down-sample the standard-exposed images (not the over or under-exposed ones) as the input LDR-LR. The HDR images corresponding to the input LDR ones are selected as the output, where they are tonemapped.

To generate reflectance dataset pairs $\{R^{i}_{LL}, R^{i}_{HH}\}$, we obtain each reflectance from the linearized luminance value. To account for the variations and non-linearities caused by non-linear camera response functions (CRFs) \cite{CRF}, the inverse of CRF is required. However, since the CRFs are usually unknown and vary diversely, we assume it just a gamma function with $\gamma < 1$ and use the inverse of CRF for the linearization. Specifically, we use $f^{-1}(x)=x^{2.2}$ for the linearization, where $f(\cdot)$ is the assumed camera response function.

It is worth pointing out that such inconsistent characteristics of HDR datasets are alleviated by removing the illumination component from the LDR images and corresponding tonemapped HDR images. We have also tried to use the original HDR image for extracting the $R$ (for removing illumination), however, we found that extraction from the tonmapped HDR image contains much better detail information, and eventually yields much better performance.

From the reflectance pairs so obtained, we extract the patches with the size of $48 \times 48$ in LR domain, which are augmented by rotation and flip. We only consider $\times 2$ for the scaling factor of super-resolution.
The mini-batch size for training is set as $32$, and the learning rate is decayed with the scale factor of $0.5$, starting from $2 \times 10^{-4}$ and halved once to $10^{-4}$. We implement the model using Tensorflow \cite{Tensorflow} library with Titan Xp GPU.

\begin{figure}[!t]
	\centering
	\includegraphics[width=0.98\columnwidth]{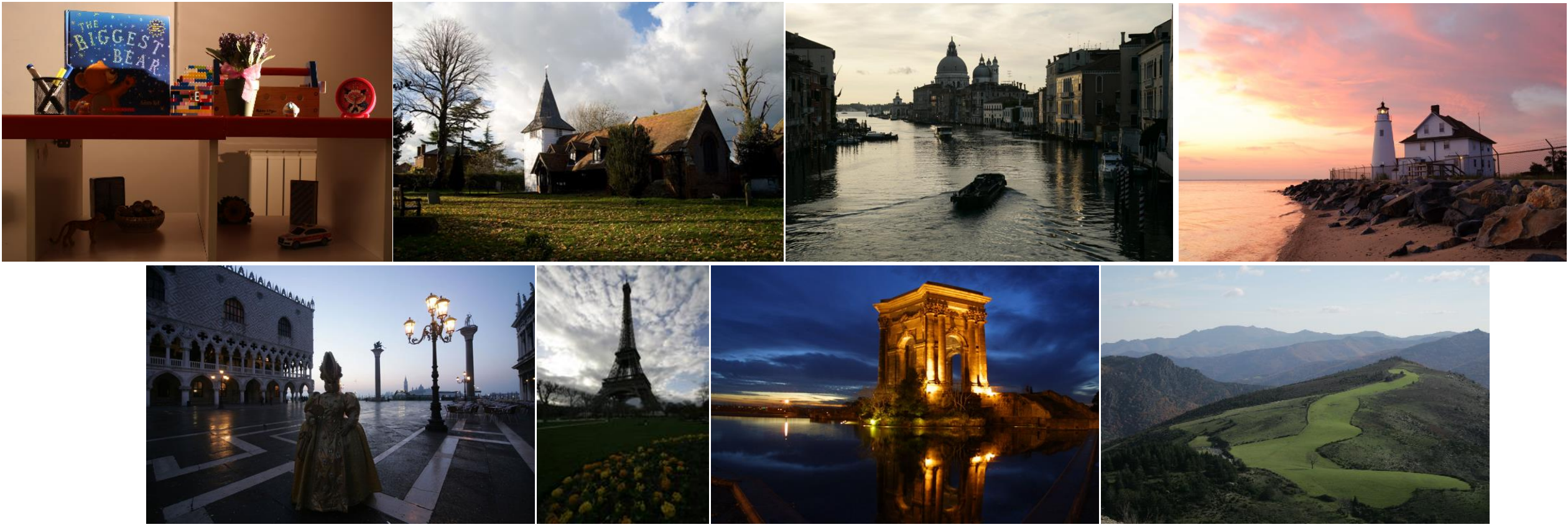}
	\caption{MESet8. From the top left, Desktoys, House, Grandcanal, Lighthouse, Mask, Tower, Peyrou, and Landscape.}
	\label{fig:MESet8}
\end{figure}

\begin{table*}[h]
	\caption{Quantitative evaluation results on the test set. The numbers are averaged metrics for the datasets. \textcolor{red}{Red} indicates the best result and \textcolor{blue}{blue} indicates the second best. In the case of HIGRADE, there are two kind of measures HIGRADE-1/HIGRADE-2.}
	\label{tab:results}
	\begin{center}
		\resizebox{\linewidth}{!}{
			\begin{tabular}{|c|c|c|c|c|c|c|} 
				\hline
				\rule[-1ex]{0pt}{3.5ex}  Metric & Dataset & LDR-LR & Kovaleski-SR & Eilertsen-SR & HDRI-SR-B (Ours) & HDRI-SR-C (Ours) \\
				\hline\hline
				\rule[-1ex]{0pt}{3.5ex} \multirow{4}{*}{NIQE($\downarrow$)} & MESet8 & 5.157 & 4.280 & 4.544 & \textcolor{blue}{4.230} & \textcolor{red}{2.746}\\
				& Part I & 5.855 & 4.700 & \textcolor{blue}{4.564} & 5.028 & \textcolor{red}{3.304}\\
				& Part II & 5.333 & \textcolor{blue}{4.558} & 4.600 & 4.649 & \textcolor{red}{3.141}\\
				& Part III & 5.415 & 4.346 & \textcolor{blue}{4.279} & 4.673 &\textcolor{red}{3.280}\\
				\hline
				\rule[-1ex]{0pt}{3.5ex} \multirow{4}{*}{HIGRADE($\uparrow$)} & MESet8 & -1.148/-0.128& -0.367/\textcolor{blue}{0.024}& -0.625/-0.107& \textcolor{blue}{-0.051}/0.015 & \textcolor{red}{0.158}/\textcolor{red}{0.287} \\
				& Part I & -1.192/-0.578 & -0.351/\textcolor{red}{0.013}&-0.657/-0.301&\textcolor{blue}{-0.253}/-0.176& \textcolor{red}{-0.104}/\textcolor{blue}{0.010}\\
				& Part II &-1.115/-0.367&-0.419/0.088&-0.610/-0.056&\textcolor{blue}{-0.030}/\textcolor{blue}{0.118}&\textcolor{red}{0.123}/\textcolor{red}{0.256}\\
				& Part III & -0.700/-0.238&\textcolor{blue}{-0.086}/\textcolor{blue}{0.205}&-0.386/-0.154&-0.109/0.135&\textcolor{red}{-0.002}/\textcolor{red}{0.266}\\
				\hline
				\rule[-1ex]{0pt}{3.5ex} \multirow{4}{*}{NQSR($\uparrow$)} & MESet8 & 5.942&\textcolor{blue}{7.015}& 6.148&6.966&\textcolor{red}{7.144}\\
				& Part I & 6.839 &\textcolor{red}{7.704}& 6.967& 6.988 & \textcolor{blue}{7.175} \\
				& Part II & 6.833 & \textcolor{red}{7.489} & 6.920 & 7.185 & \textcolor{blue}{7.346}\\
				& Part III & 6.814 & \textcolor{red}{7.312} & 6.959 & 6.900 & \textcolor{blue}{7.032}\\
				\hline
			\end{tabular}
		}
	\end{center}
\end{table*}

\section{Experimental Results}
To evaluate the proposed method, we perform the experiments on several test sets. First, we select eight well-known images from the multi-exposure images shown in \figurename{~\ref{fig:MESet8}}, which will be called the ``MESet8'' set. We also perform the experiments on three datasets called ``Part I,'' ``Part II,'' and ``Part III'' from Wang \etal's paper \cite{Wang}, which include 29, 18, and 30 images respectively.

\subsection{Comparisons}
We compare our method with the cascades of reverse tone mapping operators (rTMOs) and super-resolution (SR). As the reverse tone mapping operators, a conventional method (Kovaleski \etal's \cite{Kovaleski}) and a CNN-based method (Eilertsen \etal's \cite{HDRICNN}) are adopted. For the super-resolution algorithm, one of the recent CNN-based state-of-the-art methods, EDSR \cite{EDSR} is used. For these cascade implementations, we empirically selected the best combination, \ie, we first apply HDRI and then super-resolution because this order gives slightly better quality than its reverse order which is also evidenced in \cite{HDRSR2}.
%Also, we have tried to optimize EDSR with HDR tonemapped images, however the results were comparable as shown in the {\em supplementary material}. Therefore, we follow the original authors' implementation for EDSR.
For our proposed algorithms, two variations are demonstrated: HDRI-SR-B and HDRI-SR-C.

In summary, the following methods and input (LDR-LR) are compared.
\begin{itemize}
	\item Kovaleski-SR: HDRI by \cite{Kovaleski} followed by EDSR \cite{EDSR}.
	\item Eilertsen-SR: HDRI by \cite{HDRICNN} followed by EDSR \cite{EDSR}.
	\item HDRI-SR-B: ``Basic model,'' trained with only $\mathcal{L}_{recon}$.
	\item HDRI-SR-C: ``Complex model,'' trained with $\mathcal{L}$ based on the generative scheme.
\end{itemize}

\subsection{Metrics}
Since there is no reference image for the objective evaluation, we adopt three widely used NR-IQAs (no-reference image quality assessments). Specifically, we adopt natural image quality evaluator (NIQE) \cite{NIQE}, HDR image gradient-based evaluator (HIGRADE) \cite{HIGRADE}, and no-reference quality metric for single-image super-resolution (NQSR) \cite{NQSR}.

\begin{figure*}[t]
	\begin{center}
		\begin{subfigure}[t]{0.3\linewidth}
			\centering
			\includegraphics[width=1\columnwidth]{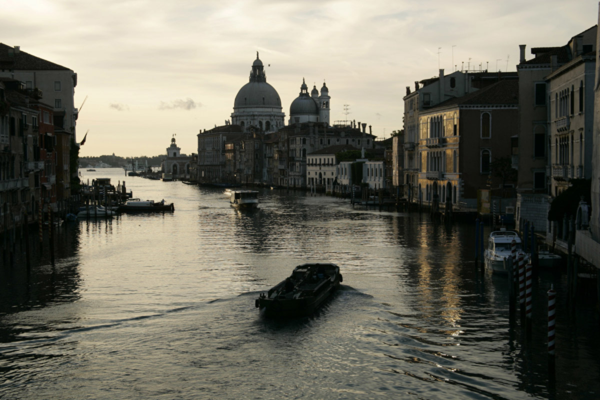}
			\caption*{LDR-LR}
		\end{subfigure}
		\begin{subfigure}[t]{0.3\linewidth}
			\centering
			\includegraphics[width=1\columnwidth]{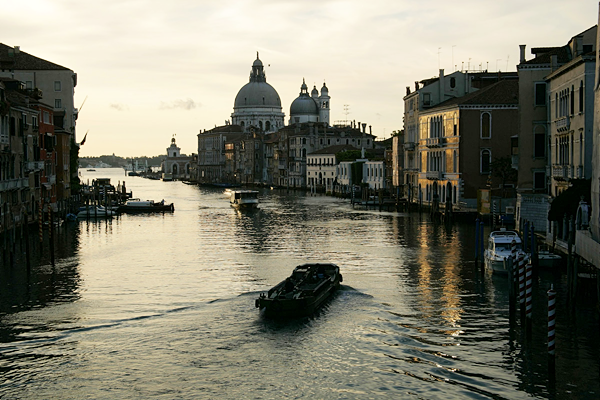}
			\caption*{Kovaleski-SR}
		\end{subfigure}
		\begin{subfigure}[t]{0.3\linewidth}
			\centering
			\includegraphics[width=1\columnwidth]{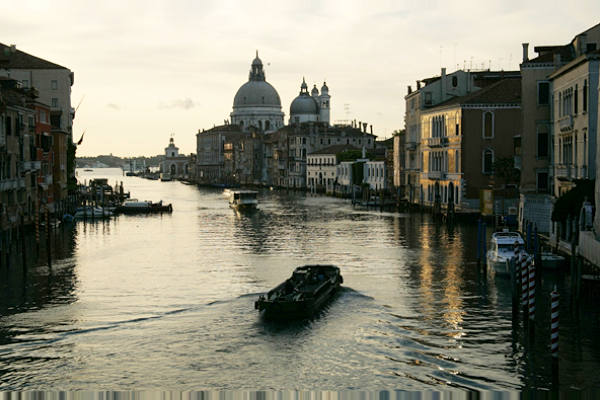}
			\caption*{Eilertsen-SR}
		\end{subfigure}
		\begin{subfigure}[t]{0.3\linewidth}
			\centering
			\includegraphics[width=1\columnwidth]{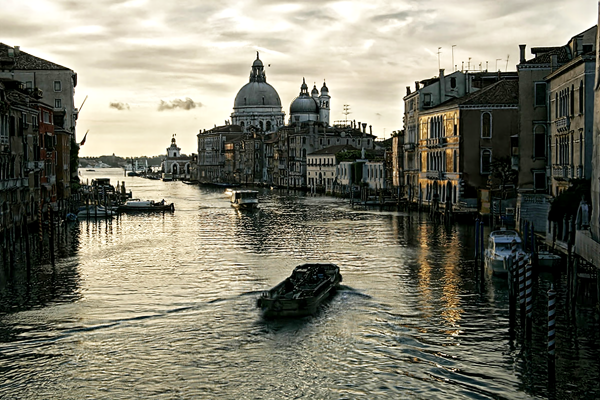}
			\caption*{HDRI-SR-B (Ours)}
		\end{subfigure}
		\begin{subfigure}[t]{0.3\linewidth}
			\centering
			\includegraphics[width=1\columnwidth]{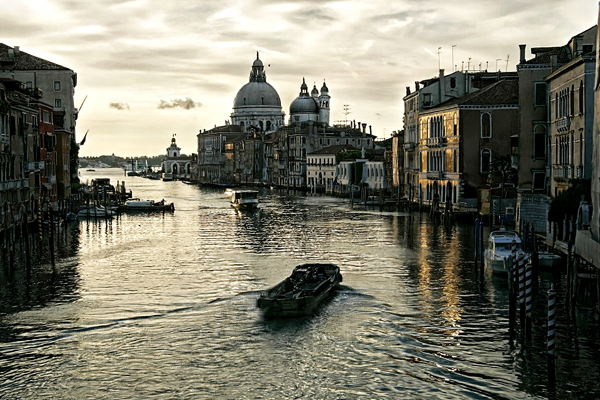}
			\caption*{HDRI-SR-C (Ours)}
		\end{subfigure}
	\end{center}
	\caption{Visualized comparisons on ``Grandcanal'' of MESet8.}
	\label{fig:002}
\end{figure*}

\begin{figure*}[t]
	\begin{center}
		\begin{subfigure}[t]{0.3\linewidth}
			\centering
			\includegraphics[width=1\columnwidth]{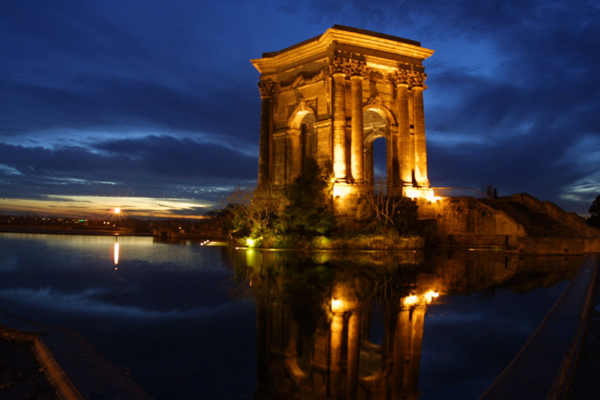}
			\caption*{LDR-LR}
		\end{subfigure}
		\begin{subfigure}[t]{0.3\linewidth}
			\centering
			\includegraphics[width=1\columnwidth]{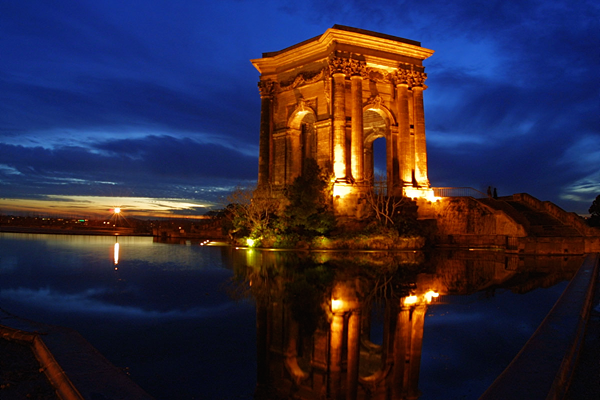}
			\caption*{Kovaleski-SR}
		\end{subfigure}
		\begin{subfigure}[t]{0.3\linewidth}
			\centering
			\includegraphics[width=1\columnwidth]{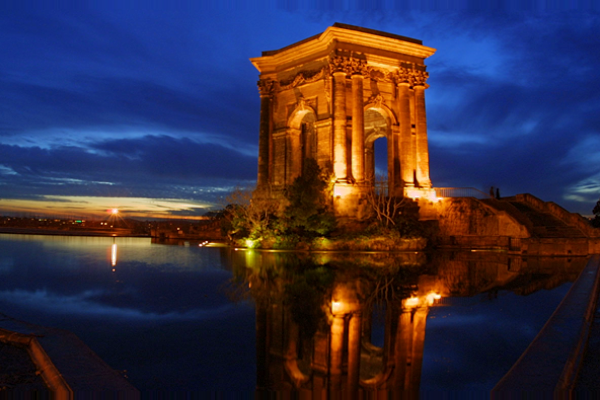}
			\caption*{Eilertsen-SR}
		\end{subfigure}
		\begin{subfigure}[t]{0.3\linewidth}
			\centering
			\includegraphics[width=1\columnwidth]{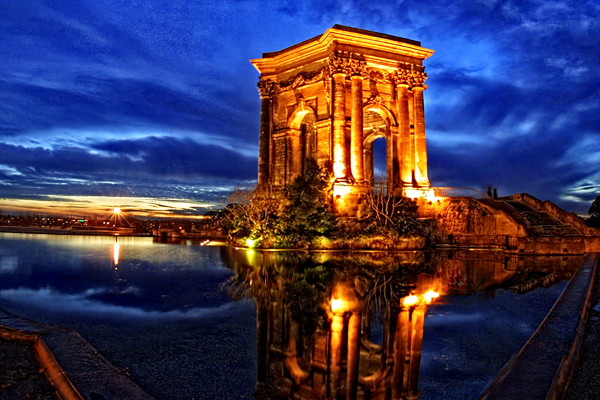}
			\caption*{HDRI-SR-B (Ours)}
		\end{subfigure}
		\begin{subfigure}[t]{0.3\linewidth}
			\centering
			\includegraphics[width=1\columnwidth]{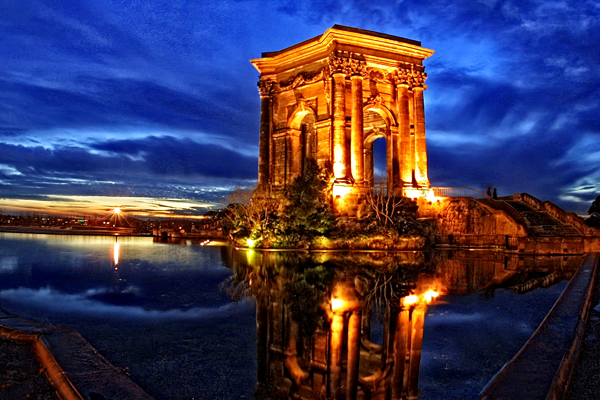}
			\caption*{HDRI-SR-C (Ours)}
		\end{subfigure}
	\end{center}
	\caption{Visualized comparisons on ``Peyrou'' of MESet8.}
	\label{fig:003}
\end{figure*}

\begin{figure*}[t]
	\begin{center}
		\begin{subfigure}[t]{0.48\linewidth}
			\centering
			\includegraphics[width=1\columnwidth]{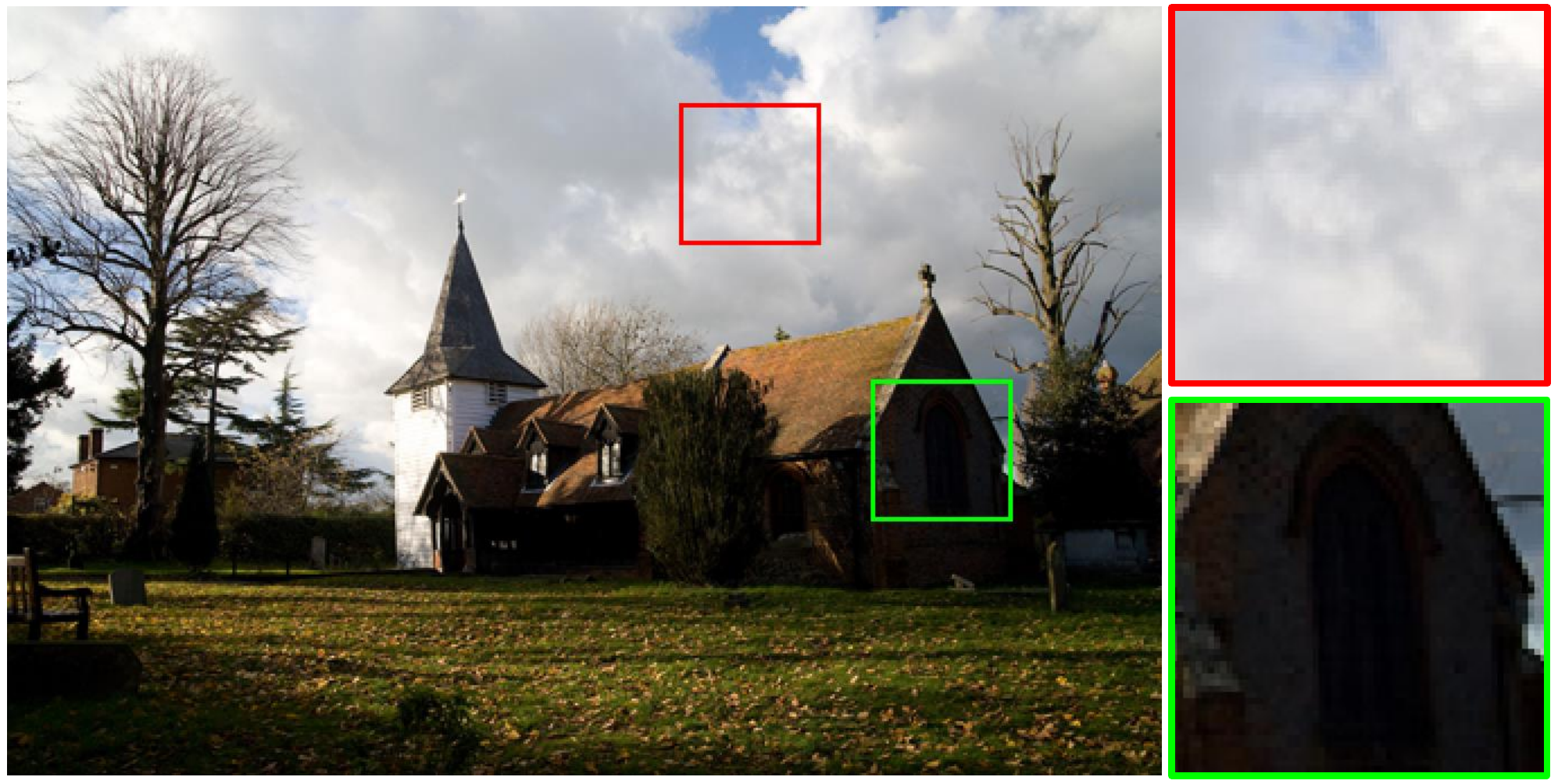}
			\caption*{LDR-LR}
		\end{subfigure}
		\begin{subfigure}[t]{0.48\linewidth}
			\centering
			\includegraphics[width=1\columnwidth]{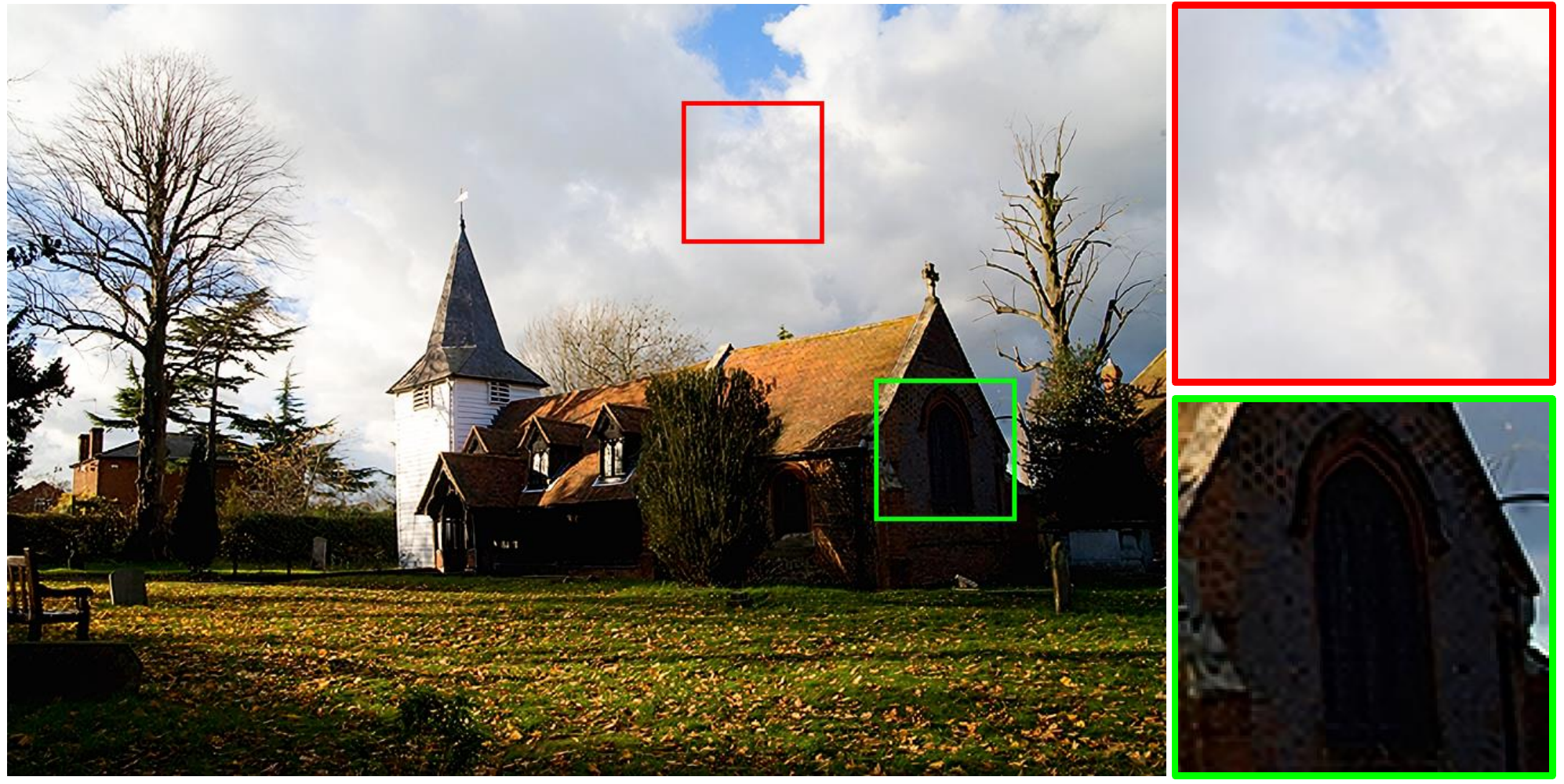}
			\caption*{Kovaleski-SR}
		\end{subfigure}
		\begin{subfigure}[t]{0.48\linewidth}
			\centering
			\includegraphics[width=1\columnwidth]{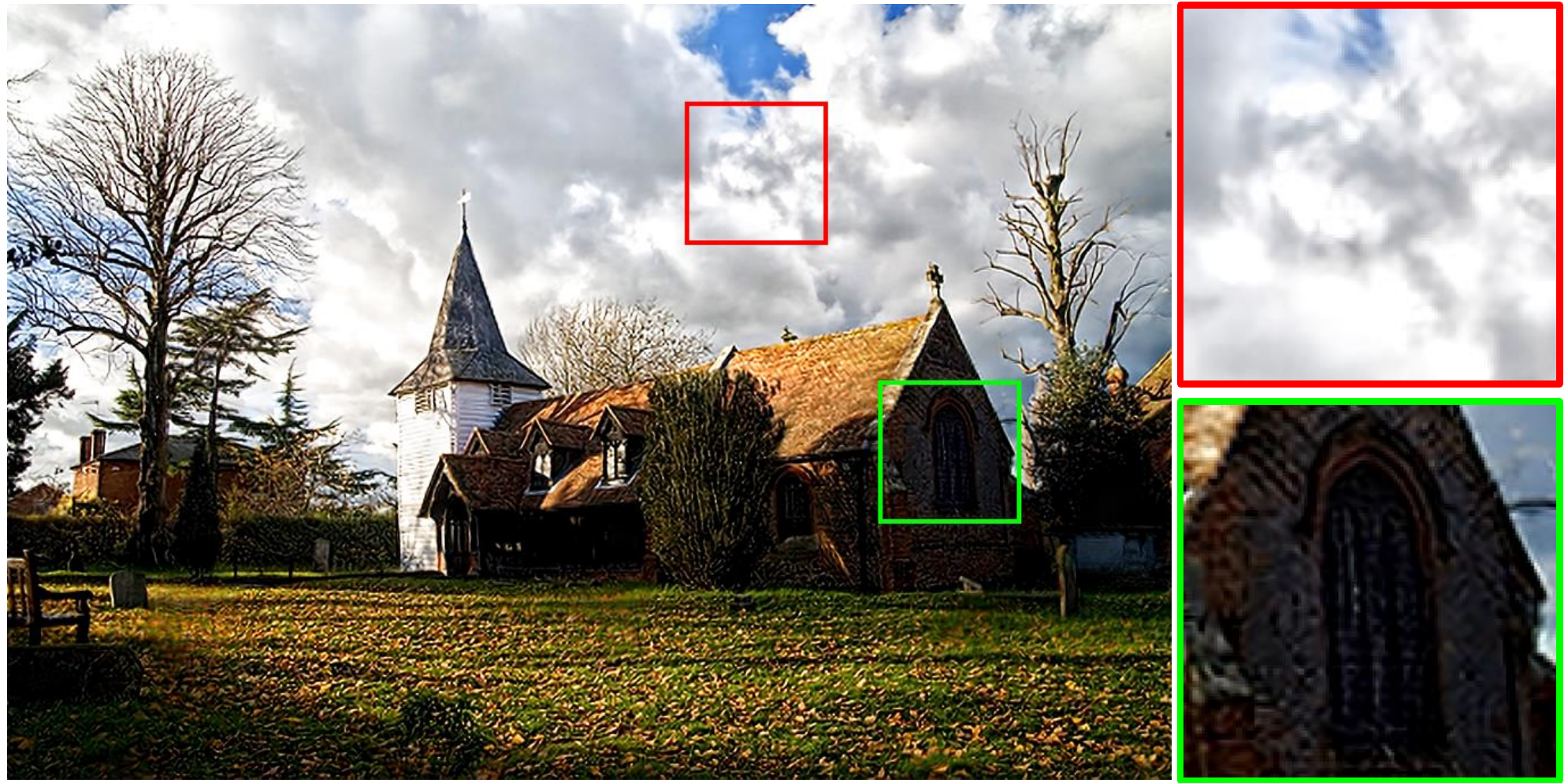}
			\caption*{HDRI-SR-B (Ours)}
		\end{subfigure}
		\begin{subfigure}[t]{0.48\linewidth}
			\centering
			\includegraphics[width=1\columnwidth]{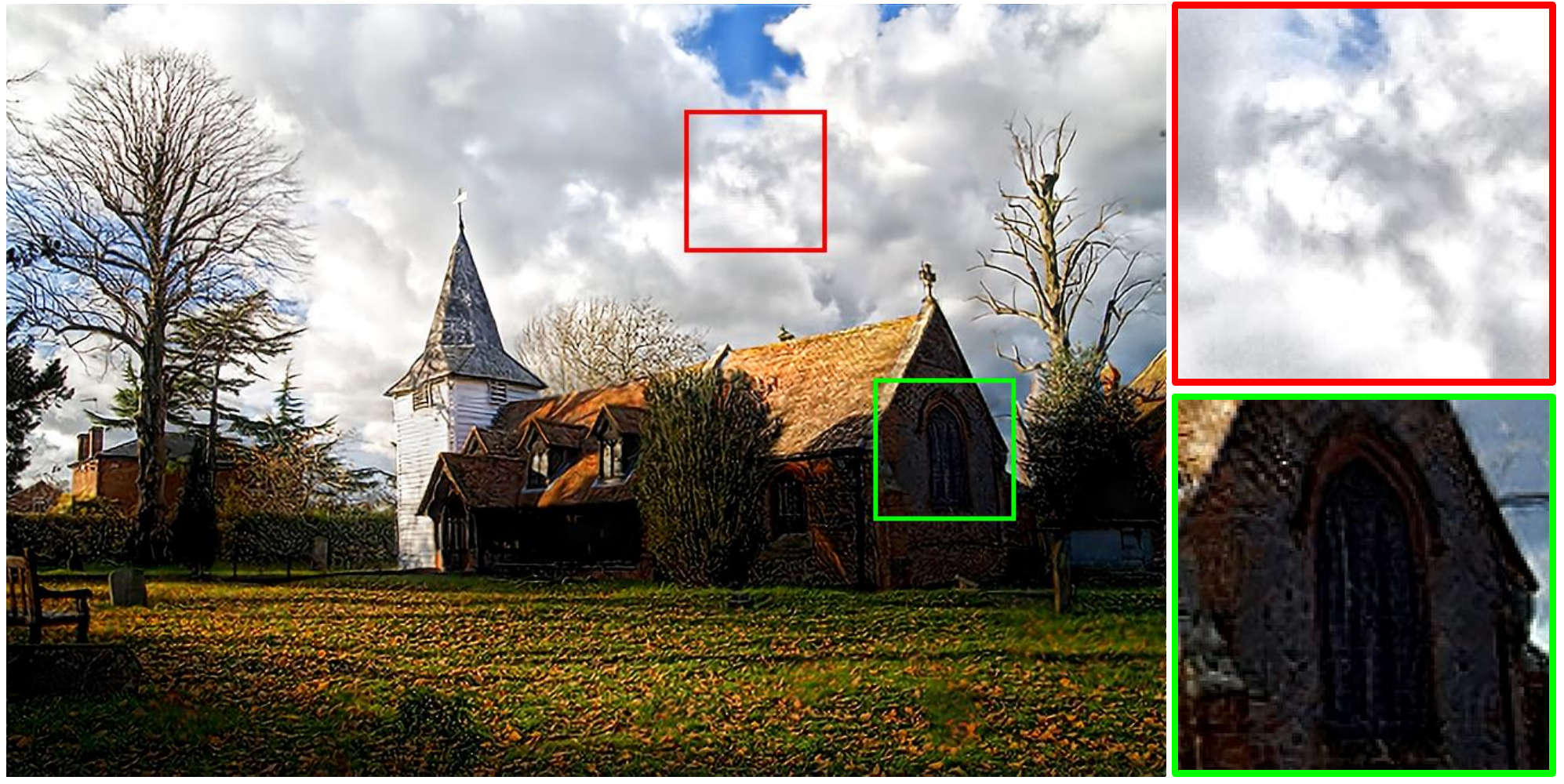}
			\caption*{HDRI-SR-C (Ours)}
		\end{subfigure}
	\end{center}
	\caption{Visualized comparisons on ``House'' of MESet8.}
	\label{fig:004}
\end{figure*}

%\begin{figure*}[t]
%	\begin{center}
%		\begin{subfigure}[t]{0.19\linewidth}
%			\centering
%			\includegraphics[width=1\columnwidth]{img/006/LDR-LR.png}
%			\caption*{LDR-LR}
%		\end{subfigure}
%		\begin{subfigure}[t]{0.19\linewidth}
%			\centering
%			\includegraphics[width=1\columnwidth]{img/006/Koval-SR.png}
%			\caption*{Kovaleski-SR}
%		\end{subfigure}
%		\begin{subfigure}[t]{0.19\linewidth}
%			\centering
%			\includegraphics[width=1\columnwidth]{img/006/Eiler-SR.png}
%			\caption*{Eilertsen-SR}
%		\end{subfigure}
%		\begin{subfigure}[t]{0.19\linewidth}
%			\centering
%			\includegraphics[width=1\columnwidth]{img/006/HDRI-SR-B.png}
%			\caption*{HDRI-SR-B (Ours)}
%		\end{subfigure}
%		\begin{subfigure}[t]{0.19\linewidth}
%			\centering
%			\includegraphics[width=1\columnwidth]{img/006/HDRI-SR-C.png}
%			\caption*{HDRI-SR-C (Ours)}
%		\end{subfigure}	
%	\end{center}
%	\caption{Visualized results on ``Tower'' of MESet8.}
%	\label{fig:006}
%\end{figure*}

\subsection{Quantitative Measurements}

\tablename{~\ref{tab:results}} shows the overall quantitative measures for all the comparisons. As expected, the input LDR-LR shows the worst result in all metrics.
In the case of NIQE \cite{NIQE} which is devised to reflect the naturalness, our complex model HDRI-SR-C shows the best results. Also, our proposed basic model HDRI-SR-B shows comparable results to the others.
The HIGRADE is designed to measure the quality of tonemapped images \cite{HIGRADE}, where two different measures HIGRADE-1 and HIGRADE-2 are defined which differ in the features they use. For these two measures, the proposed HDRI-SR-C achieves the best results, and HDRI-SR-B shows competitive results with Kovaleski-SR.
In the case of NQSR \cite{NQSR} which is designed for the quality measure of super-resolved image, Kovaleski-SR demonstrates the best result on most sets while our HDRI-SR-C achieves the second best. The HDRI-SR-C shows the best result on MESet8. As shown, our HDRI-SR-C model shows considerable super-resolution ability comparable to cascades of the HDRI and state-of-the-art EDSR.

About the computational complexity of compared methods, our CNN for the joint HDRI and SR needs 8 M parameters, Eilertsen \etal's \cite{HDRICNN} HDRI requires about 32 M, and EDSR \cite{EDSR} needs about 43 M parameters. Hence, the proposed method needs the least amount of CNN parameters among the compared ones while achieving comparable or better results as shown above.

\subsection{Visualized Evaluations}

For the qualitative evaluation, we visualize the result images by tonemapping in Figures~\ref{fig:002}, \ref{fig:003}, and \ref{fig:004}. It can be seen that ours show abundant textures and details while preserving natural tones in all figures. Also, the color and the global contrast are well enhanced with our algorithm.

Specifically in \figurename{~\ref{fig:002}}, by comparing the building facades, we can see that the high frequency details are enhanced with our HDRI-SRs. Also, the texture and the volume of water flow are much enhanced compared to other algorithms. Additionally, the volume of the cloud became much realistic.
In \figurename{~\ref{fig:003}}, we also visualized ``Peyrou'' image of MESet8. As shown, the overall contrast is enhanced within the sky and the lake. Also, the texture and details of building and the trees are enriched with ours. Additional zoomed result is shown in \figurename{~\ref{fig:001}} where it is shown that the high-frequency details and edges are enhanced and the saturated details of trees are generated.
In \figurename{~\ref{fig:004}}, visualized results of ``House'' of MESet8 are shown. For the red boxes, our algorithms show much thicker clouds than the cascade of Kovaleski \etal's \cite{Kovaleski} and the EDSR \cite{EDSR}. As HDRI-SR-C adopts the generative scheme, it generates much better cloud details compared to HDRI-SR-B. By comparing the green box regions, we can see that the diminished details due to low illumination are enhanced with our algorithms.
%More visualized results are included in the \textit{supplementary material}.
%In \figurename{~\ref{fig:006}}, visualized results of ``Tower'' of MESet8 is shown. As shown, the overall contrast and details are much enhanced. In particular, the details in sky and the volume of clouds are much more enhanced with our scheme. Also, the flowers of the front part of the image have become much realistic.
	
\section{Conclusion}
In this paper, we have proposed a CNN-based method for the joint HDRI and SR from a single image, where the domain knowledges from the existing HDRI and SR methods are exploited in designing the framework. We also considered the issue that there is no ground-truth image for the HDRI. Specifically, we found that the key to the joint task is the reconstruction of high-frequency details. 
We have also found that we can get rid of inconsistent characteristics of various HDR dataset by removing the illumination. Hence, we decompose the image into the illumination and reflectance, and process the reflectance by the CNN while we just bicubic interpolate and stretch the illumination. The final output is generated by synthesizing the processed components and then linearly stretching the synthesized image according to the target display luminance. We have also proposed a generative approach and training strategies for the joint HDRI-SR task. Experiments show that the proposed methods yield better performance than the cascade of conventional CNN-based HDRI and SR.

\pagebreak
\begin{table}[t]
	\begin{center}
		\begin{tabular}{c}
			\\[5ex]
			\fontsize{14}{0} \selectfont {\textbf{Joint High Dynamic Range Imaging and Super-Resolution from a Single Image -}}\\[0.8ex]
			\fontsize{14}{0} \selectfont {\textbf{Supplementary Material}}
		\end{tabular}
	
	\end{center}
\end{table}

\begin{figure}[h]
	\begin{center}
		\centering
		\includegraphics[width=1\columnwidth]{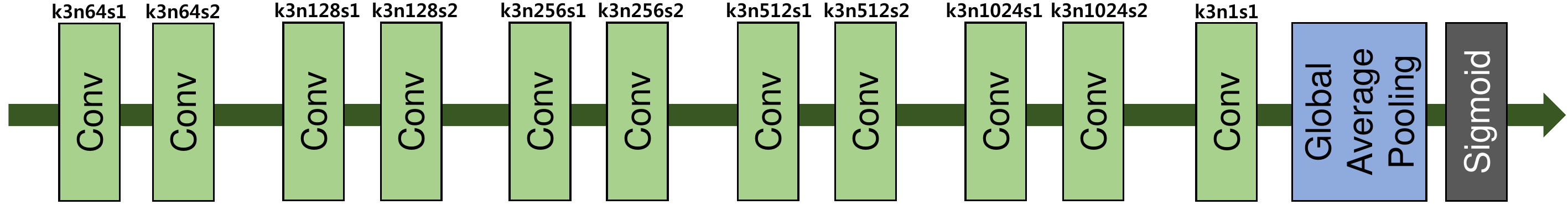}
	\end{center}
	\caption{The discriminator architecture for training HDRI-SR-C model. On each of the convolution layers, the $k, n,$ and $s$ denote kernel size, the number of features, and stride of convolution operation, respectively.}
	\label{fig:DIS}
\end{figure}

\hspace{-1.5em} {\large \textbf{1. Discriminator Architecture}}\\

\figurename{~\ref{fig:DIS} shows the discriminator architecture of our HDRI-SR-C model. We stack several convolution layers and employ global average pooling at the latter part of the network instead of fully-connected layers. Also, we alternate max-pooling operations by 2-strided convolution layers to decrease the spatial scale of features.

\end{document}